\documentclass[journal]{IEEEtran}
\usepackage{cite}

\ifCLASSINFOpdf
\else
\fi
\usepackage{url}

\usepackage{amsmath,graphicx}
\usepackage[font=small]{caption}
\usepackage{tabularx}
\usepackage{subcaption}
\usepackage{dblfloatfix}
\usepackage{fixltx2e}
\usepackage{multirow}
\usepackage{booktabs}
\usepackage{balance}

\begin{document}
%
\title{Deep Learning for Single and Multi-Session i-Vector Speaker Recognition}
%
%
%

\author{Omid~Ghahabi
        and~Javier~Hernando
\thanks{O. Ghahabi and J. Hernando are with TALP Research Center, Department of
Signal Theory and Communications, Universitat Politecnica de Catalunya - BarcelonaTech, Spain (e-mail:
omid.ghahabi@upc.edu; javier.hernando@upc.edu). 

This work has been funded by the Spanish project SpeechTech4All (TEC2012-38939-C03-02) 
and the European project CAMOMILE (PCIN-2013-067).}}
\maketitle

\begin{abstract}
The promising performance of Deep Learning (DL) in speech recognition has motivated the use of DL in other speech technology applications such as speaker recognition. 
Given i-vectors as inputs, the authors proposed an impostor selection algorithm and a universal model adaptation process in a hybrid system based on Deep Belief Networks (DBN) and Deep Neural Networks (DNN) to discriminatively model each target speaker.
In order to have more insight into the behavior of DL techniques in both single and multi-session speaker enrollment tasks, some experiments have been carried out in this paper in both scenarios. 
Additionally, the parameters of the global model, referred to as universal DBN (UDBN), are normalized before adaptation.
UDBN normalization facilitates training DNNs specifically with more than one hidden layer.
Experiments are performed on the NIST SRE 2006 corpus.
It is shown that the proposed impostor selection algorithm and UDBN adaptation process enhance the performance of
conventional DNNs 8-20\% and 16-20\% in terms of EER for the single and multi-session tasks, respectively. 
In both scenarios, the proposed architectures outperform the baseline systems obtaining up to 17\% reduction in EER.
\end{abstract}

\begin{IEEEkeywords}
Deep Neural Network, Deep Belief Network, Restricted Boltzmann Machine, i-Vector, Speaker Recognition.
\end{IEEEkeywords}

\ifCLASSOPTIONpeerreview
\begin{center} \bfseries EDICS Category: SPE-SPKR \end{center}
\fi
%
\IEEEpeerreviewmaketitle

\section{Introduction}
\label{sec:Introduction}

%
%
%
%
\IEEEPARstart{T}{he} 
recent compact representation of speech utterances known as i-vector \cite{dehak_front-end_2010}
has become the state-of-the-art in the text-independent speaker recognition.
Given speaker labels for the background data, there are also some post-processing techniques such as Probabilistic Linear Discriminant Analysis (PLDA)\cite{prince_probabilistic_2007,kenny_bayesian_2010}
to compensate speaker and session variabilities and, therefore, to increase the overall performance of the system.

On the other hand, the success of deep learning techniques in speech processing, specifically in speech recognition 
(e.g.,~\cite{mohamed_investigation_2010,dahl_context-dependent_2011,mohamed_acoustic_2011,hinton_deep_2012,AndrewSenior_icassp_2015}),
has inspired the community to make use of those techniques in speaker recognition as well.  
Three main commonly used techniques are Restricted Boltzmann Machines (RBM),
Deep Belief Networks (DBN), and Deep Neural Networks (DNN).
Different combinations of RBMs have been 
used in~\cite{stafylakis_preliminary_2012,senoussaoui_first_2012}
to classify i-vectors and in \cite{Stafylakis_RBM_PLDA_2012} to learn speaker and channel factor subspaces 
in a PLDA simulation.
RBMs and DBNs have been used
to extract a compact representation of speech signals from acoustic features \cite{vasilakakis_speaker_2013}
and i-vectors \cite{Novoselov_ODYSSEY_2014}.
RBMs have also been employed in~\cite{omid_gh_icassp_2015} as
a non-linear transformation and dimension reduction stage for GMM supervectors.
DBNs have been used in~\cite{lee_unsupervised_2009} as unsupervised
feature extractors and in \cite{pooyan_Eusipco_2015} as speaker feature classifiers. Furthermore, in~\cite{omid_gh_icassp_2014,Ghahabi_ODYSSEY_2014,Ghahabi_LNAI_2014} they have been integrated in an adaptation process to provide a better initialization for DNNs.
DNNs have been utilized to extract Baum-Welch statistics for supervector and i-vector
extraction~\cite{Campbell_Using_Deep_Belief_Networks_2014,Mclaren_icassp_2014,Kenny_ODYSSEY_2014,garcia-romero_improving_2014}.
DNN bottleneck features are recently employed in the i-vector framework \cite{Mclaren_icassp_2015,richardson_deep_2015}.
Additionally, different types of i-vectors represented by DNN architectures are proposed in \cite{variani_deep_2014,liu_deep_2015}.

The main attention of the National Institute of Standard and Technology (NIST)
over the last two years to combine i-vectors with new machine learning
techniques \cite{_nist_2014,_nist_2015} encouraged the authors to extend 
the prior works developed in~\cite{omid_gh_icassp_2014,Ghahabi_ODYSSEY_2014}. 
The authors took advantage of unsupervised learning of DBNs 
to train a global model referred to as Universal DBN (UDBN)
and DNN supervised learning to model each target speaker
discriminatively.
To provide a balanced training, an impostor selection algorithm and to cope with few training data a UDBN-adaptation process was proposed.

In this work, deep architectures with different number of layers are explored for both single and multi-session speaker enrollment tasks. The parameters of the global model are normalized before adaptation. Normalization helps to facilitate training  networks specifically where more than one hidden layer is used.
The top layer pre-training proposed in~\cite{omid_gh_icassp_2014} is not used in this work. The reason is that it emphasizes on the top layer connection weights and avoids the lower hidden layers to learn enough from the input data. This fact is of more importance where higher number of hidden layers are used. 
It is supposed, in this work, that there is no labeled background data. Moreover, no unsupervised labeling technique (e.g., \cite{Khoury_ODYSSEY_2014,Novoselov_ODYSSEY_2014}) is employed. This work shows how DNN architectures can be more efficient in this particular task.
Experimental results performed on the NIST SRE 2006 corpus \cite{_nist_2006} show that the proposed architectures  outperform the baseline systems in both single and multi-session speaker enrollment tasks.

\section{Deep Learning }
\label{sec:Deep Learning}
Deep Learning (DL) referes to a branch of machine learning techniques which attempts to learn high level features from data.
Since 2006~\cite{hinton_reducing_2006,hinton_fast_2006}, DL has become a new area of research in many 
applications of machine learning and signal processing. 
Various deep learning architectures have been used in speech processing (e.g.,~\cite{hinton_deep_2012,ling_modeling_2013,zhang_deep_2013,AndrewSenior_icassp_2015,sainath_deep_2015}).  
 Deep Neural Networks (DNN), Deep Belief networks (DBN), and Restricted Boltzmann
Machines (RBM) are three main techniques we have used in this work to discriminatively model each target speaker given input i-vectors.

\begin{figure}[t!]
        \centering
	\begin{subfigure}[t]{0.147\textwidth}
                \centering
                \includegraphics[width=\textwidth]{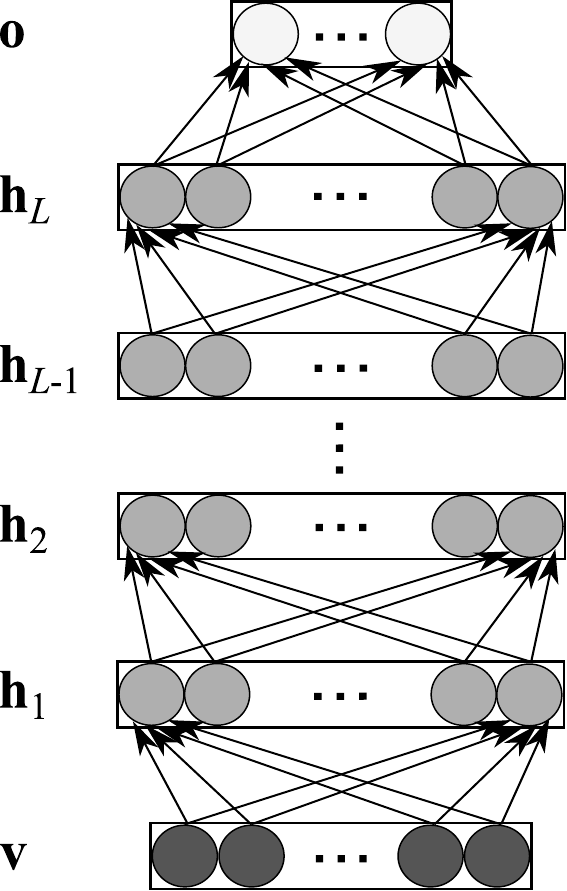}
                \caption{}
                \label{fig:DNN_IEEETrans}
        \end{subfigure}
		\quad
         \begin{subfigure}[t]{0.11\textwidth}
                \centering
                \includegraphics[width=\textwidth]{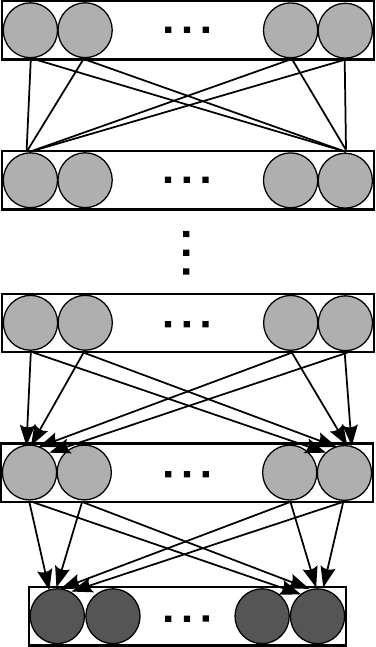}
                \caption{}
                \label{fig:DBN_IEEETrans}
        \end{subfigure}
        ~ 
        ~ 
        \begin{subfigure}[t]{0.16\textwidth}
                \centering
                \includegraphics[width=\textwidth]{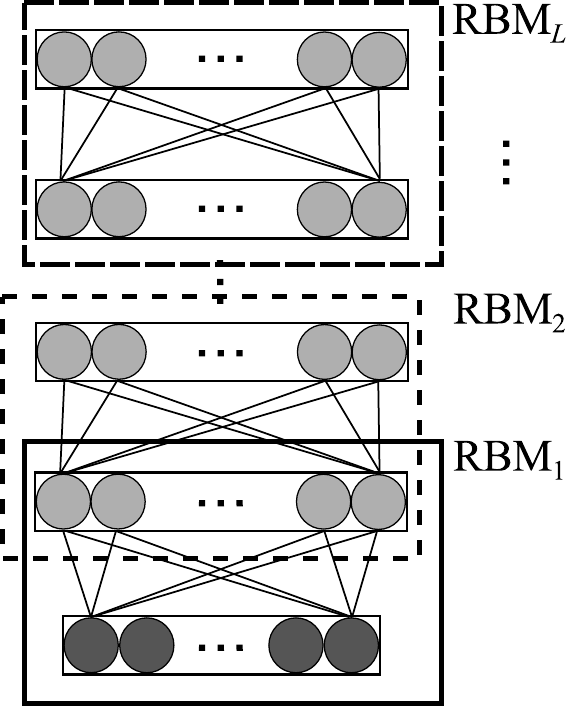}
                \caption{}
                \label{fig:DBN_Training_IEEETrans}
        \end{subfigure}
	 \caption{(a) DNN ,(b) DBN , and (c) DBN training/DNN pre-training.}
          \label{fig:DBN-DNN}
\end{figure}

DNNs are feed-forward neural networks with multiple hidden layers (Fig.~\ref{fig:DNN_IEEETrans}).
They are trained using discriminative back-propagation algorithms given class labels of input vectors.
The training algorithm tries to minimize a loss 
function between the class labels and the outputs. 
For classification tasks, cross entropy is
often used as the loss function and the softmax is commonly used as
the activation function at the output layer~\cite{ling_deep_2015}.
Typically, the parameters of DNNs are initialized with small random numbers.
Recently, it has been shown that there are more efficient techniques for parameter initialization~\cite{larochelle_exploring_2009,dumitru_erhan_difficulty_2009,erhan_why_2010}.
One of those techniques consists in initializing DNN with DBN parameters, which
it is often referred to as unsupervised pre-training or just hybrid DBN-DNN \cite{dahl_context-dependent_2011,deng_deep_2014}.
It has empirically been shown that this pre-training stage
can set the weights of the network
closer to an optimum solution than random initialization
\cite{larochelle_exploring_2009,dumitru_erhan_difficulty_2009,erhan_why_2010}. 

DBNs are generative models with multiple hidden layers of 
stochastic units above a visible layer which represents a data vector (Fig.~\ref{fig:DBN_IEEETrans}). 
The top two layers are undirected and the other layers have top-down directed connections to generate the data.
There is an efficient greedy layer wised algorithm 
to train DBN parameters \cite{hinton_fast_2006}. In this case, DBN is divided in two-layer sub-networks 
and each one is treated as an RBM (Fig.~\ref{fig:DBN_Training_IEEETrans}).
When the first RBM corresponding to the visible units is trained, its parameters are frozen and the outputs
are given to the RBM above as input vectors. This process is repeated until the top two layers are reached. 

\begin{figure}[t!]
        \centering
        \begin{subfigure}[t]{0.13\textwidth}
                \centering
                \includegraphics[width=\textwidth]{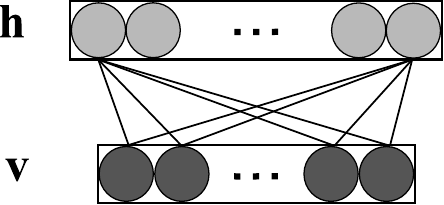}
                \caption{}
                \label{fig:RBM_IEEETrans}
        \end{subfigure}%
        \quad	
        ~ 
        \begin{subfigure}[t]{0.30\textwidth}
                \centering
                \includegraphics[width=\textwidth]{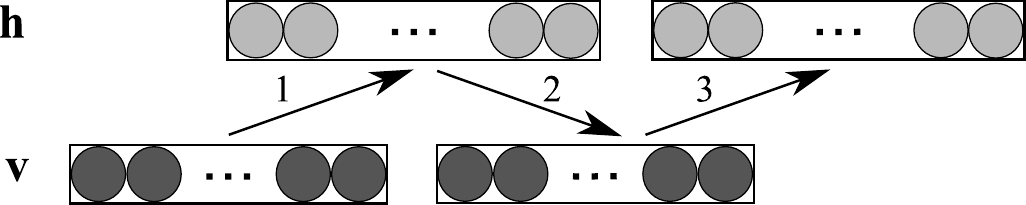}
                \caption{}
                \label{fig:RBM_Training_IEEETrans}
        \end{subfigure}
          \caption{(a) RBM and (b) RBM training.}
          \label{fig:RBM}
\end{figure}

RBMs are generative models constructed from two undirected layers of
stochastic hidden and visible units (Fig.~\ref{fig:RBM_IEEETrans}). 
RBM training is based on maximum likelihood criterion 
using the stochastic gradient descent algorithm 
 \cite{hinton_fast_2006,dahl_context-dependent_2011}.
The gradient is estimated by an approximated version 
 of the Contrastive Divergence (CD) algorithm which is called CD-1
\cite{hinton_reducing_2006,hinton_fast_2006}. As it is shown in Fig.~\ref{fig:RBM_Training_IEEETrans}, 
CD-1 consists of three steps. 
At first, hidden states (\(\textbf h\)) are computed given visible unit values (\(\textbf v\)). 
Secondly, given \(\textbf h\),
  \(\textbf v\) is reconstructed. Thirdly, hidden unit values are computed
  given the reconstructed \(\textbf v\).
Finally, the change of connection weights is given as follows,
  \begin{equation}
 \Delta_{w_{ij}} \approx  -\eta\left( \langle v_ih_j\rangle_{data}-\langle v_ih_j\rangle_{recon} \\ \right )
 \label{Eq:RBM training}
 \end{equation}
 where \(\eta\) is the learning rate, \(w_{ij}\) represents the
weight between the visible unit \(i\) and 
the hidden unit \(j\), and \(\langle .\rangle_{data}\) and \(\langle .\rangle_{recon}\) denote the
expectations 
when the hidden state values are driven from the input visible data
and the reconstructed data, respectively.
The biases are updated in a similar way.
More theoretical and practical details can be found in 
\cite{hinton_reducing_2006,hinton_fast_2006,hinton_practical_2012}. The whole training algorithm is given in~\cite{omid_gh_icassp_2015}.

In all of these networks, it is possible to update the parameters after processing each
training example,
but it is often more efficient to divide the whole input data (batch) into
smaller size batches (minibatch) and to update the parameters by averaging the gradients
over each minibatch.
The parameter updating procedure is repeated when the whole available input data
are processed. 
Each iteration is called an epoch.

\begin{figure*}[t!]
        \centering
        \includegraphics[width=0.98\textwidth]{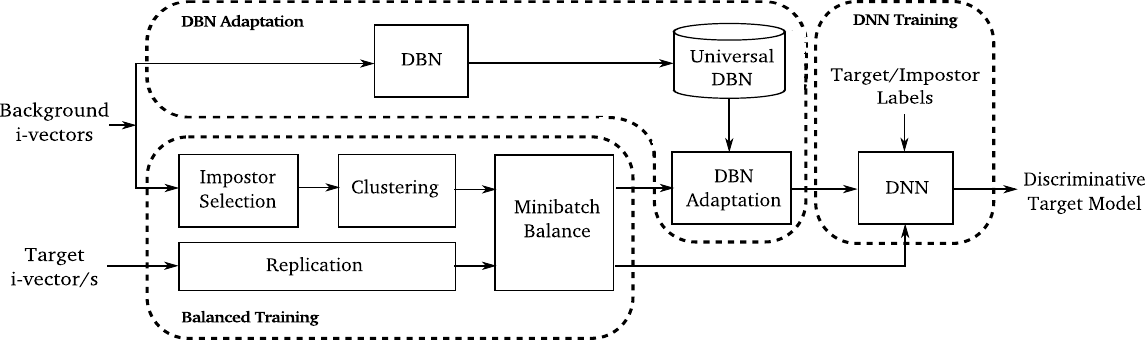}\vspace{0.50cm}
        \caption{Block-diagram of the proposed DNN based speaker recognition system.}

                \label{fig:Block-diagram}
\end{figure*}

\section{Deep Learning for i-Vectors}
\label{sec:Deep Learning for i-Vectors}
The success of the i-vector approach in speaker recognition and DL 
techniques in speech processing applications has encouraged the research community to combine those techniques for speaker recognition
\cite{omid_gh_icassp_2014,Ghahabi_ODYSSEY_2014,Ghahabi_LNAI_2014,Mclaren_icassp_2014,Kenny_ODYSSEY_2014,Mclaren_icassp_2015,richardson_unified_2015,liu_deep_2015}.
Two kinds of combination can be considered. DL techniques can be used in the i-vector extraction process \cite{Mclaren_icassp_2014,Kenny_ODYSSEY_2014,Mclaren_icassp_2015,richardson_unified_2015,liu_deep_2015},
or applied after i-vector computation
 \cite{stafylakis_preliminary_2012,senoussaoui_first_2012,Stafylakis_RBM_PLDA_2012,omid_gh_icassp_2014,Ghahabi_ODYSSEY_2014,Ghahabi_LNAI_2014}.
In order to have more insight into the behavior of DL techniques on i-vectors, in this work the authors extend the preliminary study
developed in \cite{omid_gh_icassp_2014,Ghahabi_ODYSSEY_2014}.

An i-vector \cite{dehak_front-end_2010} is a low rank vector, typically between 400 and 600, 
representing a speech utterance. 
Feature vectors of a speech signal can be modeled by a set of Gaussian Mixtures (GMM) adapted from 
a Universal Background Model (UBM).
The mean vectors of the adapted GMM are stacked to build the supervector $\mathbf m$.
The supervector can be further modeled as follows,
\begin{equation}
\small
\mathbf m =\mathbf m_{u}+\mathbf T\boldsymbol{\omega}
\label{eq:i-vector}
\end{equation}
where \(\mathbf m_u\) is the speaker- and session-independent mean supervector
typically from UBM, \(\mathbf T\) is the total variability matrix,
and \(\boldsymbol{\omega}\) is a hidden variable. 
The mean of the posterior distribution of \(\boldsymbol{\omega}\) is referred to as i-vector. 
This posterior distribution is conditioned on the Baum-Welch statistics of the given speech utterance.
The \(\mathbf T\) matrix is trained using the Expectation-Maximization (EM) algorithm given the centralized Baum-Welch statistics from background speech utterances. More details can be found in~\cite{dehak_front-end_2010}.

In the state-of-the-art speaker recognition systems, all the speech utterances, including background, train and test, 
are converted to i-vectors. Background and train i-vectors are typically called impostor and target i-vectors, respectively. 
The main objective in this work is to train a two-class DNN for each target speaker given target and
impostor i-vectors.
In the single-session target speaker enrollment task, only one i-vector is available, meanwhile in the multi-session one, several i-vectors
are available per each target speaker.
DNNs and in general neural networks usually need a large number of input samples to be trained, and
deeper networks require more input data.
The lack of enough target samples for training each DNN yields two main problems. 
Firstly, the number of target and impostor samples will be highly unbalanced, one or some few target samples 
against thousands of impostor samples. Learning from such unbalanced data 
will result in biased DNNs towards the majority class. In other words, DNNs will  
have a much higher prediction accuracy over the majority class.
Secondly, as we need to decrease the number of impostor samples
to solve the first problem, 
the total number of samples for training the network will be very few.
A network trained with such few data is highly probable to be overfitted.

Fig.~\ref{fig:Block-diagram} shows the block diagram of the proposed 
approach to discriminatively model target speakers.
In this block diagram, we propose two main contributions to tackle 
the above problems. The balanced training block attempts to decrease the number of impostor samples 
and, in the contrary, to increase the number of target ones in a reasonable and effective way. 
The most informative impostor samples for target speakers are first selected by the proposed impostor 
selection algorithm. After that, the selected impostors are clustered and the cluster centroids are considered
as final impostor samples for each target speaker model. 
Impostor centroids and target samples are then divided equally 
into minibatches to provide balanced impostor and target data in each minibatch.

On the other hand, the DBN adaptation block is proposed to compensate the lack of enough input data.
As DBN training does not need any labeled data, the whole background i-vectors are used to build 
a global model, which is referred to as Universal DBN (UDBN). The parameters of UDBN are then adapted to
the balanced data obtained for each target speaker. 
It is worth noting that as the minimum divergence training algorithm \cite{kenny_study_2008} is 
used in the i-vector extraction process, i-vectors will have a standard normal distribution \(\mathcal{N}\left(0,1
\right)\). Therefore, they will be compatible with Gaussian RBMs (GRBM) in DBN architectures,
which assume a zero-mean unit-variance normal distribution for inputs.
At the end, given target/impostor labels, adapted DBN, and balanced data, a DNN is discriminatively trained
for each target speaker. In the two following sections, we will describe these two main contributions in more details.

\section{Balanced Training}
\label{sec:Balanced Training}
As speaker models in the proposed method will be finally discriminative,
they need both positive and negative data as inputs. However, the problem is
that 
the amount of positive and negative data are highly unbalanced in this case
which yields biasing towards the majority class. 
Some of the most straightforward ways to deal with unbalanced data problem are explored in~
\cite{he_learning_2009,thai-nghe_cost-sensitive_2010,khoshgoftaar_supervised_2010}
\cite{lopez_insight_2013,barua_mwmotemajority_2014}. 
One of the simplest commonly used method is data sampling. The data of the majority class
is undersampled and, in the contrary, the data of the minority class is oversampled.
The effectiveness of these techniques is highly dependent on the data structure.

In the proposed approach in Fig.~\ref{fig:Block-diagram}, the amount of 
impostors is decreased in two steps, namely selection and clustering. On the other hand,
the amount of target samples is increased by either replication or combination.
After that, balanced target and impostor samples are distributed equally among minibatches.

\subsection{Impostor Selection and Clustering}
\label{subsec:Impostor Selection and Clustering}
The objective is to decrease the large number of negative samples in a reasonable way. 
Our proposal has two main steps. 
First, only those impostor i-vectors which are more informative 
for the training dataset are selected. 
Informative impostor means, in this case, the impostor which is not only representative to a given target
but also is close statistically to other targets in the dataset. For a real application, it makes sense to select 
those impostors who are globally close to all enrolled speakers. When the target speakers are changed,
the selected impostors can be reselected according to the new target dataset. 
Second, as the number of selected impostor samples is still high in comparison to the
number of target ones, they are clustered by the 
k-means algorithm using the cosine distance criterion. 
The centroids of the clusters are then used as the final negative samples.

\begin{figure}[t!]
        \centering
        \includegraphics[width=0.47\textwidth]{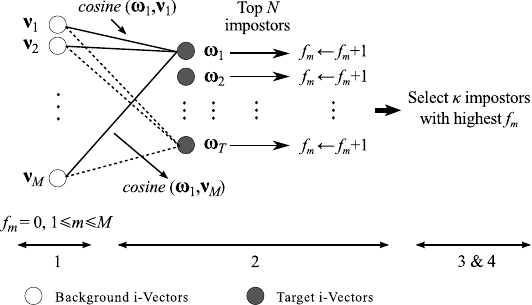}\vspace{0.50cm}
        \caption{Steps of proposed impostor selection algorithm.}

                \label{fig:Impostor_Selection_Algorithm_IEEETrans_Gray}
\end{figure}

The selection method is inspired from a data-driven background data selection
technique proposed in \cite{mclaren_data-driven_2010}.
In that technique, given all available impostors a Support Vector Machine (SVM) classifier is trained for each 
target speaker. The number of times each impostor is selected as a support vector,
in all training SVM models, is called impostor support vector frequency~\cite{mclaren_data-driven_2010}. 
Impostor examples with higher frequencies are then selected as the refined impostor dataset.
However, SVM training for each target speaker would be computationally costly. Moreover, as our final discriminative
models will be DNNs, it would not be worth to employ this technique as such.
Instead, we have proposed to use cosine distance as an efficient and a fast criterion for comparing i-vectors.
We compare each target i-vector with all impostor ones in the background dataset. Those $N$ impostors which are 
close to each target i-vector are treated like support vectors in \cite{mclaren_data-driven_2010}.
Then the $\kappa$ impostors with higher frequencies are selected as the most informative ones. The parameters $N$
and $\kappa$ are determined experimentally. The whole algorithm is shown 
in Fig.~\ref{fig:Impostor_Selection_Algorithm_IEEETrans_Gray} and can be summarized as follows,

\begin{enumerate}
   \item Set impostor frequencies \(f_m=0\), \( 1\leq m \leq M \)
  \item For each target i-vector $\boldsymbol{\omega}_{t}$, $1\leq t \leq T $
  \begin{enumerate}
    \item Compute \(cosine \left(\boldsymbol{\omega}_{t},\boldsymbol{\nu}_{m}\right)\), $1\leq m \leq M$
    \item Select the \(N\) impostors with the highest scores
    \item For the selected impostors $f_m\leftarrow f_m + 1$
  \end{enumerate}
  \item Sort impostors descendingly based on their $f_m$ 
  \item Select the first \(\kappa\) impostors as the final ones.
\end{enumerate}
where \(cosine\left(\boldsymbol{\omega}_{t},\boldsymbol{\nu}_{m}\right)\) is the cosine score
between target i-vector \(\boldsymbol{\omega}_{t}\) and the impostor i-vector $\boldsymbol{\nu}_{m}$
in the background dataset, $M$ and $T$ are the number of impostor and target speakers, respectively. 
It is worth noting that in the case of multi-session target enrollment, the average of the  
i-vectors available per each target speaker will be considered in the above selection algorithm.  

\subsection{Target Replication or Combination}
\label{subsec:Target Repeatation or Combination}
In order to balance positive and negative samples, the number of target samples
is increased as many as the number of impostor cluster centroids obtained in 
section~\ref{subsec:Impostor Selection and Clustering}.
In the single session enrollment task, the i-vector of each target speaker is simply replicated 
as many as the number of cluster centroids.
Replicated target i-vectors will not act exactly the same as each other in the pre-training process of DNNs
due to the sampling noise created in RBM training \cite{hinton_practical_2012}.
Moreover, in both adaptation and supervised
learning stages the
replicated versions make the target and impostor 
classes having the same weights 
when the network parameters are being updated. 
In multi-session task, the available i-vectors of each target speaker can be combined, i.e., 
the average of every $n$ i-vectors is considered as a new target i-vector.

\begin{figure}[t!]
        \centering
        \includegraphics[width=0.45\textwidth]{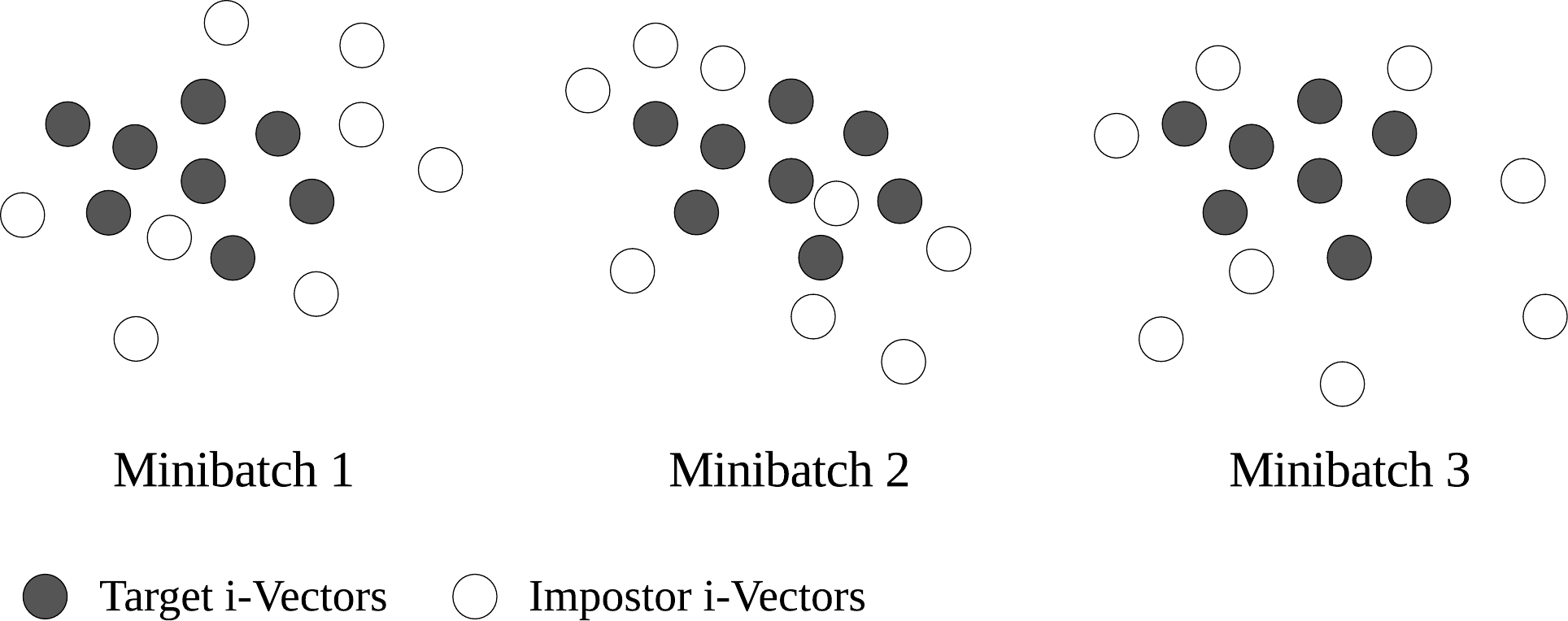}
	        \vspace{0.50cm}
        \caption{Balanced training for DNNs in multi-session speaker verification task. In each minibatch the same target i-vectors but different impostors
	are shown to DNNs.}
        \label{fig:Minibatch_Balance}
\end{figure}

Once the number of positive 
and negative samples are balanced, they are distributed equally among minibatches. 
In other words, each minibatch contains the same number of impostors and targets.
If target samples in the multi-session task are not combined, the same target samples but different impostor ones
are shown to the network in each minibatch (Fig.~\ref{fig:Minibatch_Balance}).
The optimum
numbers of impostor clusters and minibatches will be determined 
experimentally in section.~\ref{sec:Experimental_Results}.

\section{Universal DBN and Adaptation}
\label{sec:Universal DBN and Adaptation}

Unlike DNNs, which need labeled data for training, DBNs do not necessarily need
such labeled data as inputs. 
Hence, they can be used 
for unsupervised training of a global model referred to as Universal DBN (UDBN) \cite{omid_gh_icassp_2014}. 
UDBN is trained by feeding background i-vectors from different speakers. 
The training procedure is carried out layer by layer using RBMs as described in Sec.~\ref{sec:Deep Learning}.

It is shown that pre-training techniques can initialize DNNs better than simply random numbers~\cite{larochelle_exploring_2009,dumitru_erhan_difficulty_2009,erhan_why_2010}.
However, when a few numbers of input samples are available, just pre-training
may not be enough to achieve a good model.
In this case, we have proposed in \cite{omid_gh_icassp_2014} to adapt UDBN parameters to the balanced data  
obtained for each target speaker. 
Adaptation is
carried out by pre-training 
each network initialized by UDBN parameters. 
In order to avoid overfitting, only a few iterations will be used for adaptation.
It is supposed that UDBN can learn both speaker and channel variabilities from the background data.
Therefore, UDBN  will provide a more meaningful initial point for each target model than a simple random initialization.
The study in \cite{dumitru_erhan_difficulty_2009} has shown that pre-training is robust with respect to the random initialization seed.
The use of UDBN parameters makes target models almost independent from the random seeds.

In contrast to \cite{omid_gh_icassp_2014,Ghahabi_ODYSSEY_2014}, in this work we
normalize the UDBN parameters before adaptation. 
Normalization is carried out by simply scaling down the maximum absolute value of connection
weights to $0.01$. In this way, connection weights will have a dynamic range similar to that typically used for random initialization. 
Correspondingly, bias terms are multiplied by $0.01$ to be closer to zero. This is because the bias terms 
are usually set to zero when the connection weights are randomly initialized. 
UDBN parameter normalization facilitates training the networks specifically where more than one hidden layer
is used. In this way, the same learning rates and the number of epochs tuned 
for random initialized DNNs can also be used for adapted DNNs in the supervised learning stage.

\begin{figure}[t!]
        \centering
        \includegraphics[width=0.48\textwidth]{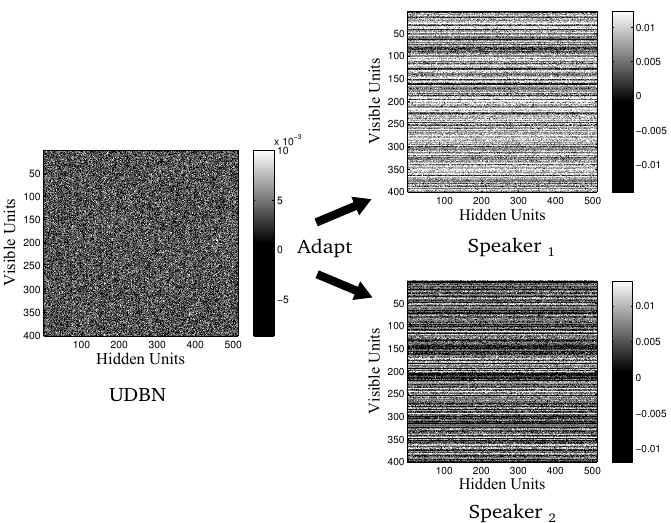}
        \caption{Comparison of the adapted connection weights between the visible and 
	  the first hidden units for two different speakers.}
        \label{fig:Initialization_Comparison}
\end{figure}

Fig.~\ref{fig:Initialization_Comparison} compares the adapted UDBN connection weights 
between the input layer and the first hidden one for two different speakers.
As it can be seen in this figure, adaptation sets speaker-specific initial points for each model.

Once the adaptation process is completed, a DNN is initialized with the adapted DBN parameters 
for each target speaker. Then the minibatch stochastic gradient descent back-propagation is carried out 
for fine-tuning. 
The softmax and the logistic sigmoid will be the activation functions of the top
label layer and the other
hidden layers, respectively.

If the input labels in the training phase are chosen as (\(1,0\)) and
(\(0,1\)) for target and impostor 
i-vectors, respectively, the final output score in the testing phase will be
computed in a Log Likelihood Ratio (LLR) form as follows,  

\begin{equation}
LLR=\log(o_1)-\log(o_2)
\label{Eq:scores}
\end{equation}
where \(o_1\) and \(o_2\) represent, respectively, the output of the first and the
second units of the top layer.
LLR computation helps to gaussianize 
the true and false score distributions which can be useful for score fusion.
In addition, to make the fine-tuning process 
more efficient a momentum factor is used to smooth out
the updates, and the
weight decay regularization is used to penalize large weights.
The top layer pre-training proposed in \cite{omid_gh_icassp_2014} is not used in this work.
The reason is that it gives the emphasis on the top layer connection weights and avoids the lower layers, closer to the input,
to learn enough from the input data. This fact will be more important when higher number of hidden layers are used.

\section{Experimental Results}
\label{sec:Experimental_Results}
The block-diagram of Fig.~\ref{fig:Block-diagram} has been implemented for both single and multi-session speaker 
verification tasks. The effectiveness of two main contributions proposed in the figure
will be shown in this section. 
\subsection{Baseline and Dataset}
\label{subsec:Baseline and Dataset}

All the experiments in this work are carried out on the NIST SRE 2006 evaluation \cite{_nist_2006}.
In both training and testing phases signals have approximately two-minute total speech duration.
The whole core condition has been used for the single session task, in which 
there are 816 target models and 51,068 trials. On the other hand, 8 conversation sides
are available per each target speaker in the multi-session task and the protocol contains
699 target models and 31,080 trials. NIST SRE 2004 and 2005 are used as the background data. 
It is worth noting that in the case of NIST 2005 only the speech signals of those
speakers who do not appear in NIST SRE 2006 are used. 

Frequency Filtering (FF) features \cite{nadeu_time_2001} are used in the experiments. FFs, like MFCCs, are 
decorrelated version of log Filter Bank Energies (FBE) \cite{nadeu_time_2001}.
It has been shown that FF features achieve a performance equal to or better than MFCCs \cite{nadeu_time_2001}.
Features are extracted every 10 ms using a 30 ms Hamming window. 
The number of static FF features is 16 and together with delta FF and delta
log energy, 33-dimensional feature vectors are produced. Before feature extraction, speech signals are subjected to an
energy-based silence removal process.

The gender-independent UBM is represented as a diagonal covariance,
512-component GMM. All the i-vectors are 400-dimensional.
The i-vector extraction process is carried out using ALIZE open source software \cite{ALIZE_3.0_2013}.
UBM and \(\textbf T\) matrix are trained using more than 6,000 speech signals
collected from NIST SRE 2004 and 2005. Performance is evaluated using 
the Detection Error Tradeoff (DET) curves, the Equal Error Rate (EER), and the minimum of the
Decision Cost Function (minDCF) calculated using \(C_M=10, C_{FA}=1 ,P_T=0.01\).

It is supposed in all experiments that there is no labeled background data and, therefore, no channel
compensation technique is used. The aim of the work is to show how DNN architectures can be more efficient in this particular task. In the baseline systems, whitened and length normalized i-vectors are classified using cosine distance.
In the multi-session task, the average of the available i-vectors per each target speaker is first length normalized and then compared with 
the test i-vector using cosine distance. In DNN experiments, raw i-vectors without whitening and length normalization are used.

\subsection{Single-Session Experiments}
\label{subsec:Single-Session Experiments}
At first, we need to choose the size of DNNs in terms of the hidden layer size and the number of layers.
The number of hidden units in each layer is taken as a power of 2 greater than the input layer size. 
Since the input layer size is 400,
the hidden layer size is chosen as 512. We explore DNNs with up to three hidden layers in all experiments. 
We do not go further than three layers because the computational complexity is increased considerably 
and also no significant gain is observed. 

As it is shown in Fig.~\ref{fig:Block-diagram},
the first step to train DNNs is to balance the number of target and impostor input i-vectors in each minibatch.
The number of minibatches and the number of impostor
centroids are set experimentally to 3 and 12, respectively.
Each minibatch will include four impostor centroids and four replicated target
samples.

After that, we train a DNN for each target speaker using
the whole impostor background data and random initialization. 
In this case, the whole background 
i-vectors are clustered using the k-means algorithm and the centroids are considered as impostor samples.
In this work, we use the uniform distribution \(\mathcal{U}\left(0,0.01 \right)\) for random initialization as the experimental results showed that it achieves slightly better performance than the normal distribution \(\mathcal{N}\left(0,0.01 \right)\) used in the prior work \cite{omid_gh_icassp_2014}. 
We tune the parameters of the networks and 
keep them fixed in all other experiments. 
One, two, and three hidden layer DNNs are trained with the learning rates of 
0.001, 0.005, and 0.08 and with the number of epochs of 30, 100, and 500, respectively.  
Momentum and weight decay are set, respectively, to 0.9 and 0.0012 for all DNNs.
In the following, we show the effect of each contribution proposed in Sec.~\ref{sec:Deep Learning for i-Vectors}.

\begin{figure}[t!]
        \centering
         \begin{subfigure}[t]{0.45\textwidth}
                \centering
                \includegraphics[width=\textwidth]{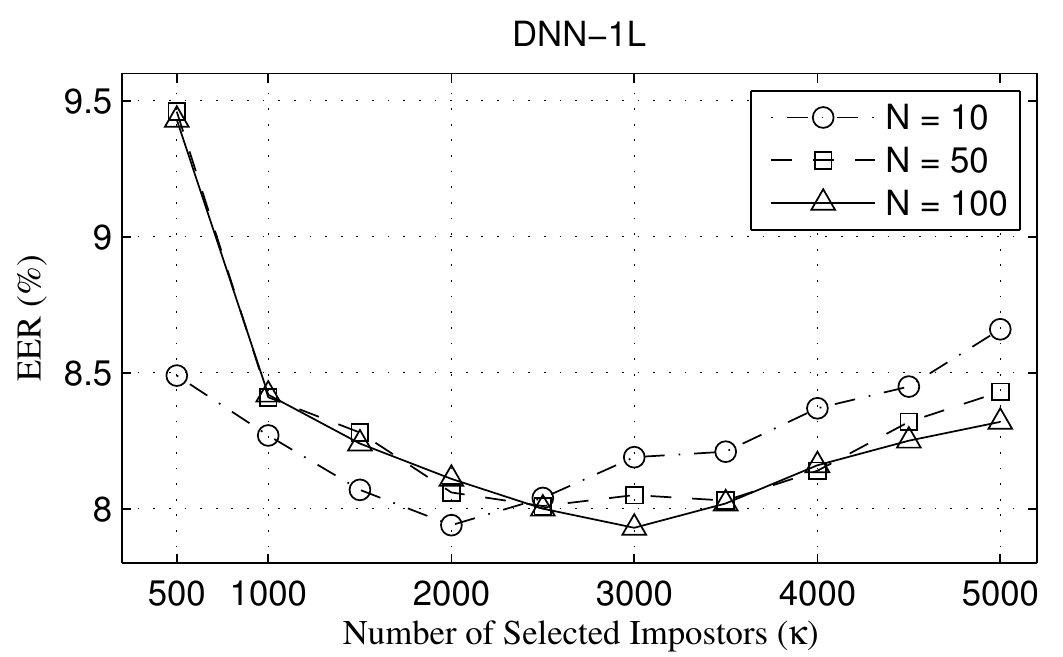}
                \caption{}
                \label{subfig:Selected_Impostors_Ntop_1LayerDNN_EER}
        \end{subfigure}
        ~ 
        ~ 
        \begin{subfigure}[t]{0.45\textwidth}
                \centering
                \includegraphics[width=\textwidth]{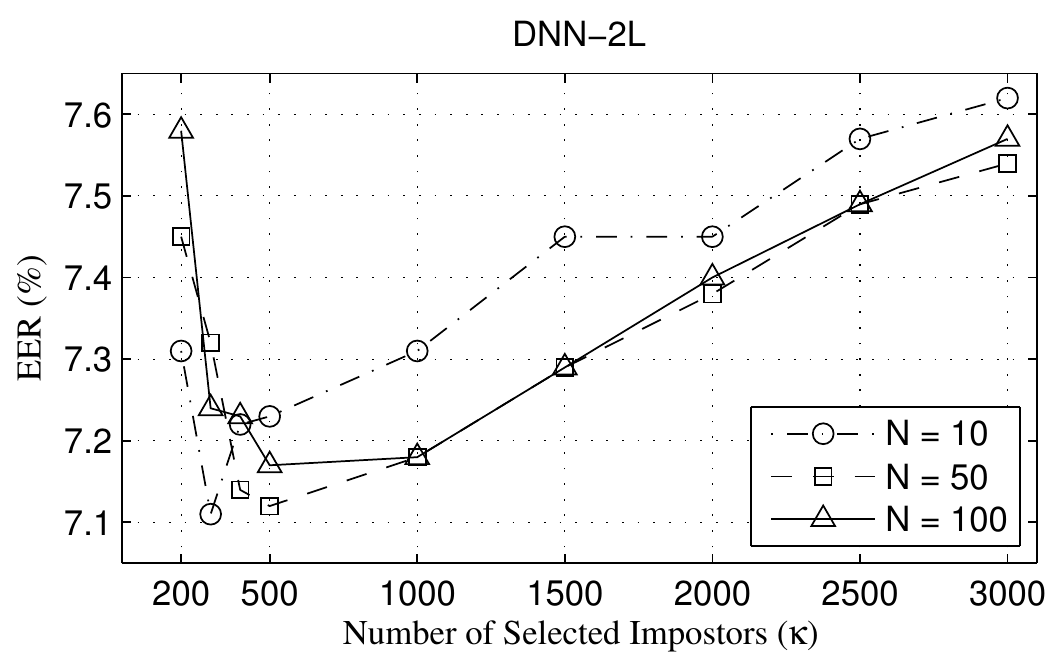}
                \caption{}
                \label{subfig:Selected_Impostors_Ntop_2LayerDNN_EER}
        \end{subfigure}
          
          \caption{Determination of the parameters of the proposed impostor selection algorithm for (a) 1 hidden layer and (b)
	    2 hidden layer DNNs. $N$ and $\kappa$ are, respectively, the number of local and global nearest impostor i-vectors
	  to the target i-vectors.}
          \label{fig:Impostor_Selection_DNN_Ntop}
\end{figure}

Background i-vectors are extracted from the same speech signals 
used for training UBM and \(\textbf T\) matrix.
The two parameters $N$ and $\kappa$, the number of local and global selected impostors in the proposed 
impostor selection method, need to be determined experimentally. 
Figures.~\ref{subfig:Selected_Impostors_Ntop_1LayerDNN_EER} and \ref{subfig:Selected_Impostors_Ntop_2LayerDNN_EER}
illustrate the variability of EER in
terms of these two parameters 
for one and two hidden layer DNNs, respectively. The same behavior
can be observed for minDCF curves. DNN with three hidden layers act similar to two-layer ones.
DNN examples shown in these two figures are initialized randomly. As it can be seen, DNNs with more than one hidden layer
tend to achieve better results with fewer number of global selected impostors in comparison to networks
with only one hidden layer. In all cases, the best performance is obtained by selecting fewer local impostors.
Hence, for all DNNs $N$ is set to 10 and $\kappa$ is set to 2,000, 300, and 500 for one, two, and 
three layer DNNs, respectively. 

UDBN is trained with the same background i-vectors used for impostor selection. 
As the input i-vectors are real-valued, a Gaussian-Bernoulli RBM (GRBM)
\cite{hinton_practical_2012,dahl_context-dependent_2011} is used to train the connection weights 
between the visible and the first hidden layer units. The rest of the connection weights are trained with
Bernoulli-Bernoulli RBMs.
The learning rate and the number of
epochs are set to 0.014 and 200 for GRBM, and to 0.06 and 120 for the rest of RBMs in UDBN, respectively. 
Momentum and weight decay are set, respectively, to 0.9 and 0.0002 for all RBMs.
Unlike in \cite{omid_gh_icassp_2014} and 
\cite{Ghahabi_ODYSSEY_2014} where the authors used the UDBN parameters as such, in this work we normalize the 
connection weights so that the maximum absolute value is $0.01$.  
This is the maximum value of the random numbers we used to initialize DNNs.
Additionally, we scale down the bias terms by $0.01$. Normalization helps to facilitate 
training DNNs with higher number of layers. Moreover, we can use the same learning rates 
we tuned for random initialized DNNs.

Experimental results showed that the most part of the improvement due to the adaptation process
comes from the adaptation of the 
connection weights between the input layer and the first hidden layer for all DNNs. The adaptation of the other 
layers has no positive effect or it improves the performance slightly. We adapted the networks up to two layers.
The learning rate of adaptation is set to 0.001 and 0.0001 for the first and the second layers, respectively.
The number of epochs for the first layer is set to 10, 20, and 15 for DNN-1L, -2L, and -3L, respectively.
DNN-1L stands for a one hidden layer DNN.
The number of epochs for the second layer is set, respectively, to 15 and 20 for DNN-2L and DNN-3L. 

Table~\ref{table:table_single_2006} summarizes the effect of each proposed contribution.
In the first row of the table, DNNs are initialized randomly and the impostor cluster centroids are obtained on the whole background data. 
As it can be seen in this row, adding more hidden layers to the network improves the performance. 
However, they still work worse than
the baseline system in which i-vectors are classified using cosine distance. 
EER and minDCF for the baseline system are 7.18\% and 0.0324, respectively.
Impostor selection improves the performance to a great extent
for all the networks. However, the biggest improvement due to the adaptation process is observed in DNNs with one 
hidden layer. The best results are obtained using both impostor selection and adaptation techniques which show 
an 8-20\% and 10-17\% relative improvements in terms of EER and minDCF, respectively, in comparison to 
the conventional DNNs. 
The last row of the table shows 
the fusion of DNN systems with the baseline in the score level. Scores of each system are first 
mean and variance normalized and then summed. 
Although DNNs with one hidden layer yield slightly better results, 
DNNs with more layers provide complementary information to the baseline system. This confirms the theoretical
hypothesis which states that more hidden layers more abstractions from the input data. 
The fusion of baseline and DNNs with three hidden
layers achieves the best results corresponding to an 8\% relative improvement for both EER and minDCF in comparison to the baseline system. We have also combined the scores of DNNs with different number of hidden layers, but no gain is observed. 

\begin{figure}[t!]
        \centering
        \includegraphics[width=0.45\textwidth]{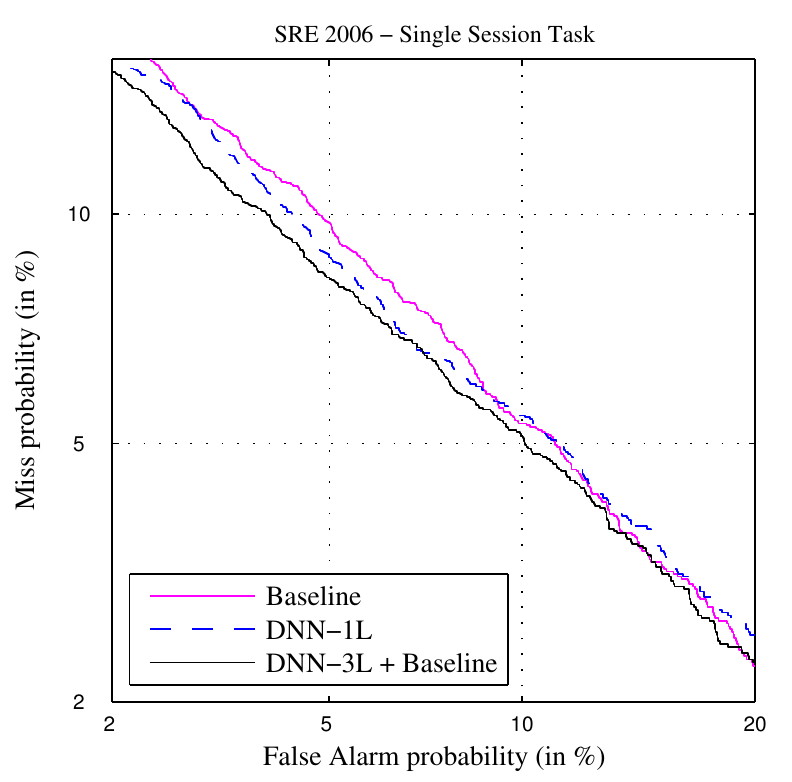}
        \caption{Comparison of the performance of the proposed DNN based systems with the baseline system (i-vector + cosine). DET curves are obtained on the core test condition of NIST SRE 2006.}
        \label{fig:DET_Curve_Single_Session}
\end{figure}

The DET curve in Fig.~\ref{fig:DET_Curve_Single_Session} 
compares the best systems in all operating points. As it is shown in this figure, DNNs with one hidden layer achieve
better results than the baseline and the combination of 3-layer DNNs with the baseline works 
the best in all operating points.

\begin{table} [t!]
\centering
\small 
\caption{{The effect of each proposed idea of Fig.~\ref{fig:Block-diagram} on the performance of the DNN target models. Results are obtained 
on the core test condition of NIST SRE 2006. Baseline is classification of i-vectors using cosine distance with EER=7.18 and minDCF=324.}}
\vspace{-1mm}
\setlength{\tabcolsep}{5pt}
\renewcommand{\arraystretch}{0.5}
\begin{tabular}{@{}c c c c c c c c @{}}
\toprule
\multirow{4}{*}{\begin{minipage}{0.54in}\textbf{Impostor Selection}\end{minipage} } & 
\multirow{4}{*}{\textbf{Adaptation}} & \multicolumn{3}{c}{\textbf{EER} (\%)} & 
\multicolumn{3}{c}{\textbf{minDCF} $(\times 10^4)$} \\\cmidrule(l){3-8}
         &                    & \multicolumn{3}{c}{ \# Hidden Layers} & \multicolumn{3}{c}{\# Hidden Layers} \\\cmidrule(l){3-8}
                  &                   &     1  &    2   &    3  &   1    &   2    &   3   \\ 
\midrule
\midrule
No & No  &  8.55 & 7.76 & 7.59 & 381 & 353 & 351\\
\midrule
Yes & No  & 8.06  & 7.12 & 7.09 & 360  & 327 & 326\\
\midrule
No & Yes  & 7.43 & 7.47 & 7.45 & 339 & 343 & 339\\
\midrule
Yes & Yes  &  \textbf{6.81} & 6.97 & 6.99 & 315  & 317 & 313\\
\midrule
\multicolumn{2}{l}{ \textbf{Fusion with Baseline}} & 6.83 & \textbf{6.88} & \textbf{6.64} & \textbf{308} & \textbf{309} & \textbf{299}\\
\bottomrule
\end{tabular}
\label{table:table_single_2006}
\end{table}

\subsection{Multi-Session Experiments}
\label{subsec:Multi-Session Experiments}
The same configuration used for the single session task is also applied for the multi-session one.
The number of minibatches is set to 3. In each minibatch, all 8 target i-vectors accompanying with 
8 impostor cluster centroids are shown to the network. Therefore, the size of each minibatch and the 
total number of impostor clusters will be 16 and 24, respectively.
As the combination of the i-vectors of each target speaker did not help the training of the networks, 
we replicated the target i-vectors in every minibatch as it was shown in Fig.~\ref{fig:Minibatch_Balance}. 

We start training the networks with the same parameters tuned for the single session experiments.
However, as the target i-vector samples per each training speaker are different from each other 
in the multi-session task, the number of epochs and the learning rates need to be slightly re-tuned.
We have set the learning rates to 0.001, 0.01, and 0.08 and the number of epochs to 
30, 100, and 500 for one layer to three layer DNNs, respectively. 

Results are summarized in Table~\ref{table:table_multi_2006}. 
Around 12\% relative improvements are achieved in all DNNs employing impostor selection technique proposed in this work.
With the same parameters obtained for the single session task, we re-selected the impostors for the new multi-session
data set.
The adaptation process improves 
the performance up to 8\%. As in the single session task, adaptation is more effective in the one hidden layer DNNs. 
For all the networks, only the parameters of the first hidden layer are adapted 
because no more improvement was observed adapting the other layers. 
Adaptation is carried out by the learning rate of 0.001 for all DNNs and the 
number of epochs of 10, 10, and 25 for DNNs with one to three layers, respectively.
The best results are obtained with three layer DNNs when the two proposed techniques are used together.
It shows more than 20\% improvement of EER and minDCF in comparison
to the conventional three layer DNNs. 
Compared to the baseline system in which EER and minDCF are obtained 4.2\% and 0.0191, respectively, 
the proposed three hidden layer DNNs achieve more than 
17\% and 10\% improvements in terms of EER and minDCF, respectively.
Fusion with the baseline system at the score level improves the results in all cases. Fusion is effective
mostly on the minDCF which increase the improvement from 10\% to 15\%.

\begin{figure}[t!]
        \centering
        \includegraphics[width=0.45\textwidth]{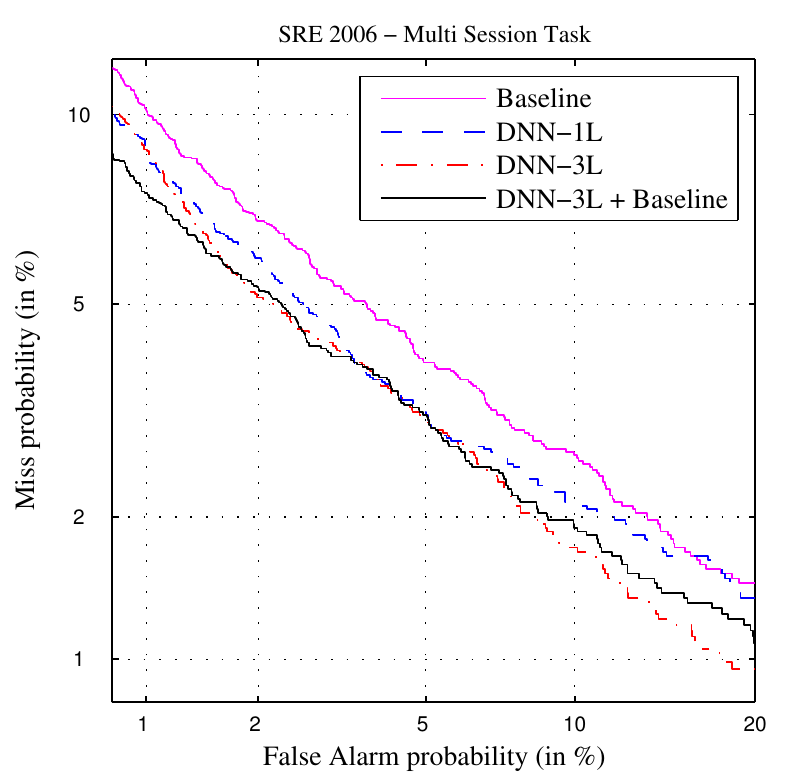}
        \caption{Comparison of the performance of the proposed DNN based systems with the baseline system (i-vector + cosine). DET curves are obtained on the 8-session enrollment task of NIST SRE 2006.}
        \label{fig:DET_Curve_Multi_Session}
\end{figure}

Fig.~\ref{fig:DET_Curve_Multi_Session} compares the DET curves of the best results obtained in
table~\ref{table:table_multi_2006}. As it can be seen in this figure, DNN-3L outperforms 
clearly the baseline and the DNN-1L 
in all operating points. However, fusion with the baseline system improves the performance only 
for the operating points with higher false alarm probabilities.

\section{Conclusion}
\label{sec:Conclusion}
A hybrid system based on Deep Belief Networks (DBN) and Deep Neural Networks (DNN) has been proposed in this work for 
speaker recognition to discriminatively model target speakers with available i-vectors.
In order to have more insight into the behavior of these techniques in both single and multi-session speaker enrollment tasks, the experiments are carried out in both scenarios. 
Two main contributions have been proposed to make DNNs more efficient in this particular task.
Firstly, the most informative 
impostors have been selected and clustered to provide a balanced training. Secondly, 
each DNN has been initialized with the speaker specific parameters adapted from a 
global model, which has been referred to as Universal DBN (UDBN).
The parameters of UDBN are normalized before adaptation, which facilitates the training of DNNs 
specifically with more than one hidden layer.
Experiments are performed on the NIST SRE 2006 corpus. 
It was shown that in both scenarios the proposed architectures outperform the baseline systems with up to 17\% and 10\% in
terms of EER and minDCF, respectively. 

\begin{table} [t!]
\small 
\caption{{The effect of each proposed idea of Fig.~\ref{fig:Block-diagram} on the performance of the DNN target models. Results are obtained 
on NIST SRE 2006, 8-session enrollment task. Baseline is classification of i-vectors using cosine distance with EER=4.2
and minDCF=191.}}
\vspace{-1mm}
\setlength{\tabcolsep}{5pt}
\renewcommand{\arraystretch}{0.5}
\centerline{
\begin{tabular}{@{} c c c c c c c c @{}}
\toprule
\multirow{4}{*}{\begin{minipage}{0.54in}\textbf{Impostor Selection}\end{minipage} } & 
\multirow{4}{*}{\textbf{Adaptation}} & \multicolumn{3}{c}{\textbf{EER} (\%)} & 
\multicolumn{3}{c}{\textbf{minDCF} $(\times 10^4)$} \\\cmidrule{3-8}
         &                    & \multicolumn{3}{c}{ \# Hidden Layers} & \multicolumn{3}{c}{\# Hidden Layers}\\\cmidrule{3-8}
                  &                   &     1  &    2   &    3  &   1    &   2    &   3   \\ 
\midrule
\midrule
No & No  &  4.58 & 4.58 & 4.38 & 208 & 213 & 217\\
\midrule
Yes & No  & 4.02  & 4.07 & 3.86 & 183  & 201 & 194\\
\midrule
No & Yes  & 4.24 & 4.30 & 4.20 & 202 & 207 & 202\\
\midrule
Yes & Yes  &  3.68 & 3.83 & 3.50 & 170  & 189 & 172\\
\midrule
\multicolumn{2}{l}{ \textbf{Fusion with Baseline}} & \textbf{3.61} & \textbf{3.77} & \textbf{3.45} & \textbf{161} & \textbf{169} & \textbf{162}\\
\bottomrule
\end{tabular}}
\label{table:table_multi_2006}
\end{table}


%




\ifCLASSOPTIONcaptionsoff
  \newpage
\fi



%



\bibliographystyle{IEEEbib}
\bibliography{Mybib}

%


\begin{IEEEbiography}[{\includegraphics[width=1in,height=1.25in,clip,keepaspectratio]{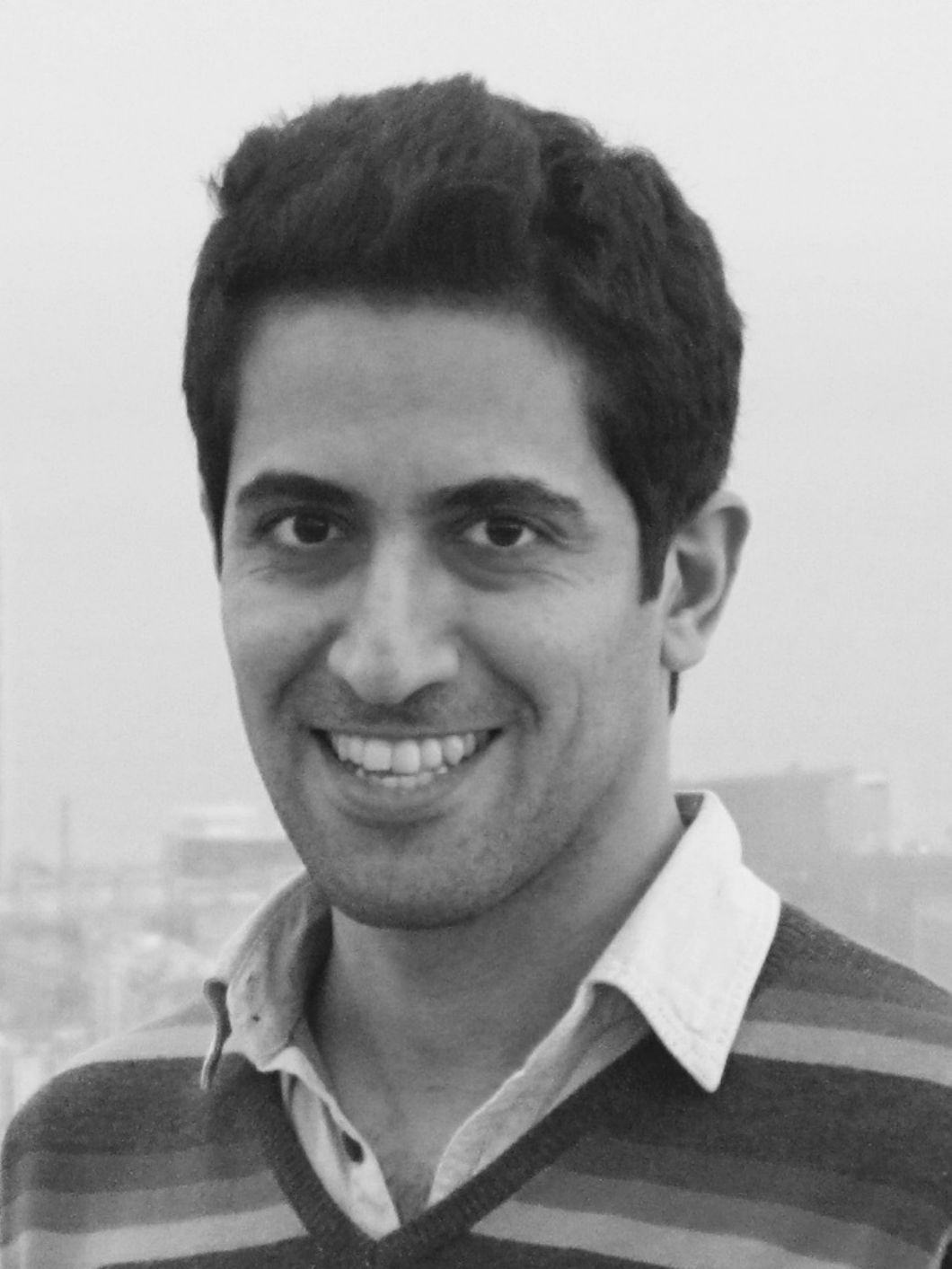}}]{Omid Ghahabi}
received the M.Sc. Degree in electrical engineering from Shahid
Beheshti University, Tehran, Iran, in 2009. From 2009 to 2011, 
he has been with the speech processing group of the Research 
Center for Intelligent Signal Processing (RCISP), Tehran, Iran.
He is now a Ph.D. candidate at Technical University of Catalonia (UPC)-BarcelonaTech, Spain. 
Since 2011, he has been working as a researcher in the speech processing group of the Signal Theory and
Communications Department of UPC. He is also a member of the Research Center for 
Language and Speech Technologies and Applications (TALP), Barcelona, Spain. His 
research interests include, but not limited to, speaker recognition, speech 
signal processing, and deep learning. He is the author and coauthor of several journal 
and conference papers on these topics. 
\end{IEEEbiography}
\begin{IEEEbiography}[{\includegraphics[width=1in,height=1.25in,clip,keepaspectratio]{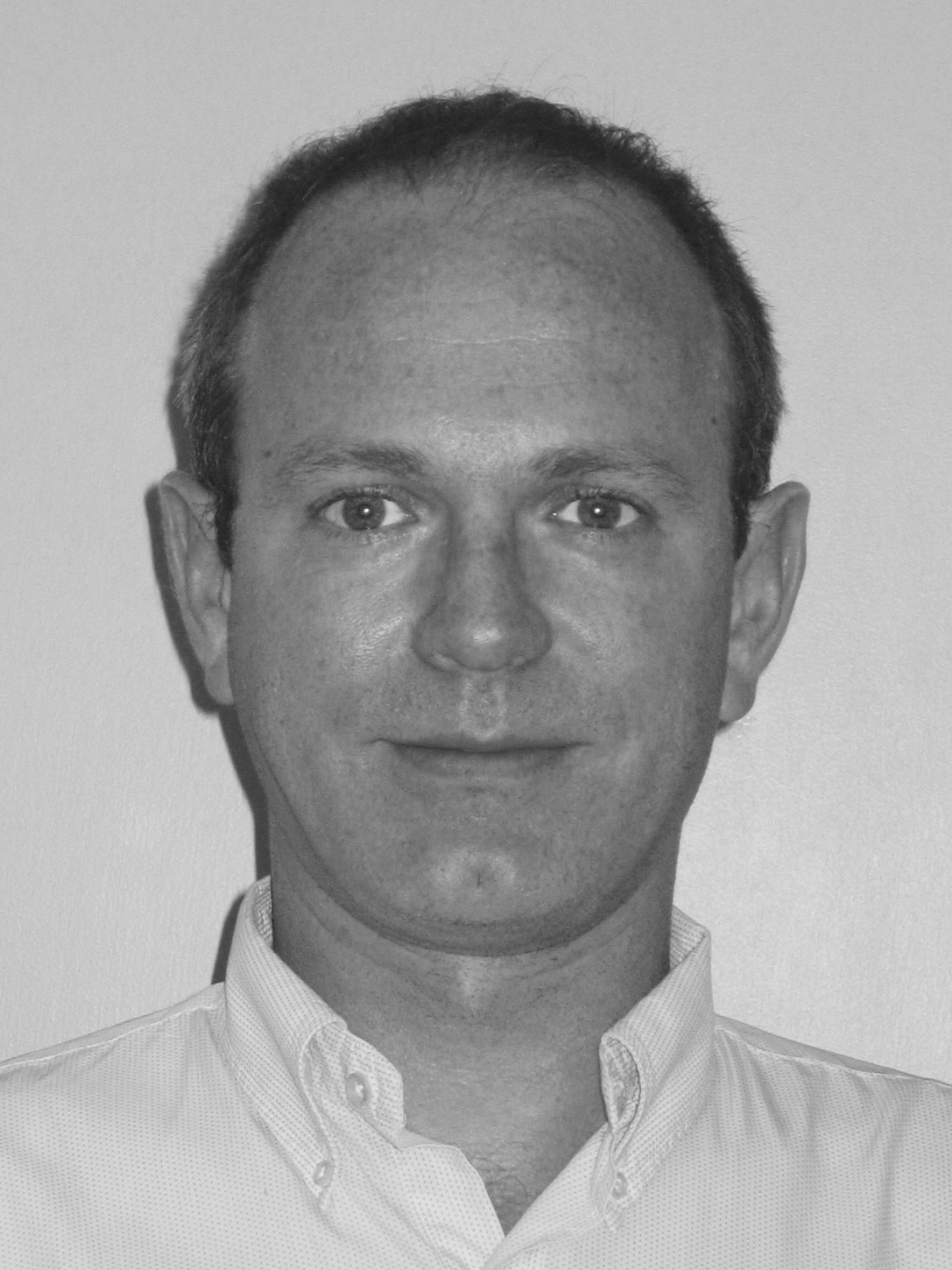}}]{Javier Hernando}
received the M.S. and Ph.D.
degrees in telecommunication engineering from the
Technical University of Catalonia (UPC), Barcelona,
Spain, in 1988 and 1993, respectively. Since 1988,
he has been with the Department of Signal Theory
and Communications, UPC, where he is a Professor
and a member of the Research Center for Language
and Speech (TALP). He was a Visiting Researcher at
the Panasonic Speech Technology Laboratory, Santa
Barbara, CA, during the academic year 2002-2003.
His research interests include robust speech analysis,
speech recognition, speaker verification and localization, oral dialogue, and
multimodal interfaces. He is the author or coauthor of about two hundred
publications in book chapters, review articles, and conference papers on these
topics. He has led the UPC team in several European, Spanish and Catalan
projects. Prof. Hernando received the 1993 Extraordinary Ph.D. Award of UPC.
\end{IEEEbiography}
%
\vfill
%



\end{document}